\documentstyle[preprint,aps,eqsecnum]{revtex}

\begin{document}

\tightenlines
\draft

\title{Freezing and clustering transitions for penetrable spheres}

\author{C. N. Likos,$^{1}$ M. Watzlawek,$^{2}$ and H. L\"owen$^{1,2}$}
\address {$^{1}$ Insitut f\"ur Festk\"orperforschung, Forschungszentrum
J\"ulich GmbH, D-52425 J\"ulich, Germany\\
$^{2}$ Institut f\"ur Theoretische Physik II, Heinrich-Heine-Universit\"at
D\"usseldorf,\\ Universit\"atsstra{\ss}e 1, D-40225 D\"usseldorf, Germany
\\ ~\hfill~ \\ ~\hfill~ \\}

\date{published in Phys. Rev. E {\bf 58}, 3135 (1998)}
     
\maketitle

\begin{abstract}
We consider a system of spherical particles interacting by 
means of a pair potential equal to a finite constant 
for interparticle distances smaller than the sphere diameter and zero outside.
The model may be a prototype for the interaction between micelles
in a solvent [C. Marquest and T. A. Witten, J. Phys. France {\bf 50},
1267 (1989)].
The phase diagram of these penetrable spheres is investigated 
using a combination of cell- and density functional theory for the
solid phase together with simulations for the fluid phase. The
system displays unusual phase behavior due to the fact that,
in the solid, the optimal configuration is achieved when certain
fractions of lattice sites are occupied by more than one particle,
a property that we call `clustering'. We find that freezing from 
the fluid is followed, by increasing density, by a cascade of
second-order, clustering transitions in the crystal.
\end{abstract}

\narrowtext

\section{Introduction}

Much of our current understanding of the liquid-solid transition from a
microscopic point of view is based on the density-functional theory
of inhomogeneous liquids \cite{evans,singh,loewen}.
This approach allows, in principle,
the systematic
calculation of the phase diagram of any system, once the
pair potential between its constituent particles is given. A number
of pair interactions of variable `hardness' (hard spheres, inverse-power,
Yukawa etc.) have been studied, yielding the phase coexistence between
a fluid phase, which is stable up to moderate densities,
and a crystal which is stable at higher densities.
For most of the systems which have been considered in the
literature,
the assumed pair
interaction between particles has the property that it grows as
the distance between the particles decreases, and diverges
at zero separation. These are the usual, {\it unbounded} interactions.
For such interactions, a whole mechanism of liquid-state 
integral equation theories has been developed which allows one
to calculate with a high degree of accuracy the structure and
thermodynamics of the fluid phase, which is in turn a necessary
ingredient in any density-functional treatment of the freezing
transition.

Much less is known about interactions which are {\it bounded}, i.e.\
they allow the particles to `sit on top of each other', imposing
only a finite energy cost for a full overlap. This is natural since
a true, microscopic interaction always forbids overlaps. However,
the situation may be different if, e.g., one considers the 
`potential of mean force' between two polymeric coil centroids 
in a good solvent, as suggested many years ago by 
Stillinger \cite{still}. The two centroids may coincide without  
this resulting into a forbidden configuration. Stillinger thus introduced
the `Gaussian core model', consisting of particles that interact by means 
of a pair potential $\phi(r) = \phi_0 \exp(-r^2/\sigma^2)$, 
where is $r$ is the
interparticle distance, $\phi_0$ is an energy scale and
$\sigma$ is a length scale. This
model and its phase diagram have been examined in Refs.\
\onlinecite{still,stillweb}, following an approach based on 
general mathematical properties particular to the Gaussian potential
and on computer simulations, for a review see Ref.\ \onlinecite{stillrev}.

In this paper we also consider a bounded potential, albeit an
apparently simpler one. We take an interaction between spheres 
which is simply equal to some positive constant if there is any
overlap between them and zero otherwise. The study of such a model
is not of purely academic interest; a few years ago, Marquest 
and Witten \cite{marquest} suggested that interaction potentials
qualitatively similar to a step function are expected for 
micelles in a solvent. 
We study the
phase diagram of this model by using standard techniques (integral
equation theories for the fluid and a cell model for the solid),
also combined with computer simulations. We find, on the
one hand, that the boundedness of the interaction makes the 
standard integral equation theories inadequate to accurately
describe the dense liquid phase of the system. On the other hand,
the fact that the interaction is constant, brings about a novel
possibility for the crystal to lower its free energy, namely the
formation of groups of two or more particles (`clusters') occupying the same
lattice site, a property that we call {\it clustering}. As a result,
there are second-order clustering transitions within the region
of the phase diagram occupied by the solid.

The rest of the paper is organized as follows: in section II we
present our approach for the fluid phase and in section III for
the solid phases. The results are combined in section IV where we
present the phase diagram of the model. Finally, in section V
we summarize and conclude.

\section{Penetrable sphere model: the fluid phase}

We consider a model of penetrable spheres, 
whose interactions are described by the pair potential:
\begin{eqnarray}
\phi(r)=\cases{\varepsilon& $\;\;\;$ $0 \leq r < \sigma$;\cr
               0          & $\;\;\;$ $ \sigma < r$,}
\label{pot1}
\end{eqnarray}
where $\sigma$ is the diameter of the spheres and $\varepsilon$ is the 
height of the energy barrier ($\varepsilon > 0$). The packing fraction
$\eta$ and reduced temperature $t$ are defined as:
\begin{eqnarray}
\eta = {\pi\over{6}}\rho\sigma^3;\;\;\;\;\;\;t = {k_{B}T\over{\varepsilon}},
\label{therm}
\end{eqnarray}
where $\rho$ is the number density, $T$ is the temperature 
and $k_{B}$ is Boltzmann's constant.

Clearly, at zero temperature
the model reduces to the hard sphere (HS) potential. The first task is
to investigate the structure and thermodynamics of the fluid state.
In a theoretical approach to the problem, typically one of the various
approximate liquid-state integral equation theories is employed, which
yields the radial distribution function $g(r)$ of the fluid together
with the direct correlation function $c(r)$ related to $g(r)$ by means
of the Ornstein-Zernicke (OZ) relation\cite{hansen}:
\begin{eqnarray}
g(r) - 1 = c(r) + \rho\int c(|{\bf r} - {\bf r'}|)[g(r') - 1] d{\bf r'}.
\label{oz}
\end{eqnarray}   

Another 
exact relation connecting $g(r)$ with $c(r)$ reads as:
\begin{eqnarray}
g(r) = \exp\{-\beta\phi(r)+g(r)-1-c(r)-B(r)\},
\label{exact} 
\end{eqnarray}
where $B(r)$ is the so-called bridge function\cite{mhnc}, the sum of
all elementary diagrams that are not nodal. Since $B(r)$ is not known,
the various approximate liquid-state integral equation theories can
be regarded as approximations of this quantity. In this way, an
additional equation or `closure' involving only $g(r)$ and $c(r)$ is
supplemented to the OZ-relation and the system becomes solvable.

The simplest and most frequently employed theories
are the Hypernetted Chain (HNC) and Percus-Yevick (PY) schemes which,
however, due to their approximate character lack thermodynamic
consistency; the `pressure' and `compressibility' routes to the
liquid free energy yield different results.
In the HNC, one 
simply sets $B(r) = 0$, obtaining the closure:
\begin{eqnarray}
g(r) = \exp\{-\beta\phi(r)+g(r)-1-c(r)\}.
\label{hnc}
\end{eqnarray}
On the other hand, the Percus-Yevick closure can be seen as a linearized
version of the HNC regarding the term $g(r)-1-c(r)$ in the exponential
and reads as:
\begin{eqnarray}
g(r) = e^{-\beta\phi(r)}[g(r)-c(r)],
\label{py}
\end{eqnarray}
corresponding to the following approximation for the bridge function:
\begin{eqnarray}
B_{{\rm PY}}(r) = [g(r)-c(r)] - 1 - \ln[g(r)-c(r)].
\label{brpy}
\end{eqnarray}

There have been various attempts to improve the above approximations and to
come up with a manageable theory which would also overcome the problem
of thermodynamic inconsistency mentioned above. Among the most popular
are the Modified HNC (MHNC) approach of 
Rosenfeld and Ashcroft \cite{mhnc} and the theory of Rogers and Young
(RY) \cite{ry}. In the latter, one replaces the exact relation 
(\ref{exact}) above by the closure:
\begin{eqnarray}
g(r) = \exp\{-\beta\phi(r)\}
\Biggl[1 + {{\exp\{\gamma(r)f(r)\}-1}\over{f(r)}}\Biggr],
\label{ryclos}
\end{eqnarray}
where $\gamma(r) = g(r) - c(r) - 1$ and $f(r)$ is a `mixing function'
depending on a single parameter $\zeta$ and taken to have the form:
\begin{eqnarray}
f(r) = 1 - \exp(-\zeta r).
\label{mixing}
\end{eqnarray}
The parameter $\zeta$ is determined in such a way that thermodynamic
consistency is achieved. The nomenclature `mixing function' comes from
the fact that the RY-closure provides a means of interpolation between
the PY- and HNC-closures.

In order to obtain a comparison and test the performance of integral
equations, we have also performed standard Monte-Carlo simulations
\cite{allen} in the constant $NVT$-ensemble. All runs were performed
in a cubic box containing 500 particles and using periodic boundary
conditions. We calculate the radial distribution function $g(r)$
as well as the structure factor $S(k)$ `on the flight'. 

For $t = 0$, where our model reduces to hard spheres, the PY solution
is analytic and is known to describe the pair structure of the HS fluid
quite well. As a first step, therefore, we have solved the PY closure
for finite temperatures as well. In Fig.\ \ref{compgr} we show results 
for $g(r)$ and 
in Fig.\ \ref{compsk} for the structure factor $S(k)$ for $t = 0.2$ and 
packing fraction $\eta = 0.3$
in comparison with simulation. In Fig.\ \ref{comppy} we compare
the $g(r)$'s for the same temperature 
and $\eta = 0.5$. 
As can be seen, for the lower density, $g(r)$ is
reproduced quite well by the PY-closure outside the core. However,
inside the core the simulation shows a tendency of $g(r)$ to grow
towards the origin, which is {\it not} reproduced by the PY-result.
The growth of $g(r)$ towards the origin can be simply understood as
follows: since the interaction is such that it does not impose 
any additional penalty for full sphere overlaps (in comparison 
with partial ones), as the density grows there is an increasing
tendency of the particles to form clusters in which more and more
spheres `sit on top of each other'. In this way, more space is left
free for the remaining clusters and the optimal configuration is
achieved. The discrepancies between the PY and the true results are
not dramatic for $\eta {< \atop \sim} 0.3$ and this limit grows 
with decreasing temperature.
Moreover, the discrepancies in the structure factor are much less
pronounced than those for the distribution function.
However, the differences
become really spectacular as the packing fraction grows. The PY-closure
is inadequate to reproduce the accumulation of spheres on top of
each other and brings about a radial distribution function that is
quite wrong at high densities.

The failure of the PY-closure to describe the very dense liquid at 
finite temperatures is not a surprise; after all, it is known that
PY works best for hard, short-range interactions like hard spheres.
Thus, we resorted to the HNC as a possible solution. 
In Fig.\ \ref{compgr} we show the comparison of the HNC-$g(r)$ with
simulation for the data point $t=0.2$, $\eta = 0.3$. As can be seen, 
now the penetration towards the origin is {\it overestimated}. In fact,
this feature becomes more and more pronounced as $\eta$ grows and,
as a result, the HNC fails to converge any more for 
$\eta {> \atop \sim} 0.6$
at $t = 0.1$. We can qualitatively understand the overestimation of 
$g(r)$ inside the core by the HNC as follows: it is well-known that
the bridge function is a positive-definite quantity \cite{mhnc}
(although this has
not been strictly proven, it turns out to be true in almost all cases)
and thus it plays the role of an `effective repulsive interaction'.
Then the HNC, by setting $B(r) = 0$ everywhere, gives rise to a $g(r)$
which is too high. For the case of an unbounded interaction which
diverges at the origin, inaccuracies in the approximation of $B(r)$
especially for low $r$ where this function is relatively large,
cause no serious problems. Indeed, 
referring to Eq.\
(\ref{exact}) we see that if $\phi(r) \to \infty$
as $r \to 0$, then the interaction dominates in the exponential
and sends $g(r) \to 0$ for short separations. But in our case where
$\phi(r)$ remains finite for all $r$, an accurate knowledge of the 
bridge function is {\it essential} in order to bring about a sensible
theory for this system.

The Rogers-Young closure provides a more sophisticated
approximation for $B(r)$.
We have attempted, therefore, to solve this closure but again we ran
into difficulties: no self-consistent solution could be found for
$\eta {> \atop \sim} 0.45$ for $t = 0.1$. Moreover, the 
results for the lower values of $\eta$ were very similar to the PY-ones.
Further attempts to modify and improve the RY closure did not yield
the desired agreement with the simulations. We do not expect that 
any other of the standard closures will be of much use either, for
the reasons described above: in the formulation of 
all approximate liquid-state theories
it is assumed (explicitly or implicitly) that the 
strongly repulsive interaction simply forbids close approaches between
particles, so that there exists some 
(generally temperature- and density-dependent)
distance $r_0$, such that 
for $r < r_0$ the radial distribution function $g(r)$
vanishes.  
Here, the situation is quite the opposite: the 
interaction is such that it {\it favors} close approaches (in fact:
full overlaps) at high density. Thus, we have decided to resort 
entirely to computer simulations in order to calculate the structure
and thermodynamics of the fluid phase at high densities. 

There are two ways or `routes' to evaluate the excess free energy of
a fluid from a simulation. The density route or `$\eta$-route'
consists of performing
a series of simulations at {\it fixed} temperature but for increasingly
high densities. Once the radial distribution function $g(r;\eta)$ has
been calculated, the form of the interaction at hand implies that
the excess pressure is related to the `jump' of $g(r)$ by the equation:
\begin{eqnarray}
{{\beta P_{ex}}\over{\rho}} = 
 4 \eta [g(\sigma^{+};\eta)-g(\sigma^{-};\eta)],
\label{jump}
\end{eqnarray}
where $g(\sigma^{\pm};\eta)$ is the value of $g(r)$ immediately 
outside/inside the core. Then, the excess free energy per particle 
is obtained by:
\begin{eqnarray}
{{\beta F_{ex}(\eta)}\over{N}} = \int_{0}^{\eta}
{{\beta P_{ex}}\over{\rho}} {{d \eta'}\over{\eta'}}.
\label{denroute}
\end{eqnarray} 

Another way to calculate the excess free energy is by the so-called
temperature route or `$t$-route'. Here, one makes use of the thermodynamic
identity relating the excess energy per particle, 
$U_{ex}/N \equiv u_{ex}(\beta)$
and the reduced excess free energy per particle, $\beta F_{ex}/N \equiv
\beta f_{ex}(\beta)$ at {\it fixed} density:
\begin{eqnarray}
u_{ex}(\beta) = {{\partial [\beta f_{ex}(\beta)]}\over{\partial \beta}},
\label{thermo}
\end{eqnarray}
to express the latter as an integral from $T = \infty$ (where
$u_{ex}$ vanishes) to the considered value of the temperature:
\begin{eqnarray}
\beta f_{ex}(\beta) = \int_0^{\beta} u_{ex}(\beta') d \beta'.
\label{troute}
\end{eqnarray}
Thereby, a series of simulations is performed at fixed $\eta$ but at 
successively decreasing temperatures. For each temperature, the value
of the internal energy is measured and at the end the integral of 
Eq.\ (\ref{troute}) is performed. Notice that for the interaction at
hand, the evaluation of the internal energy in a simulation is 
particularly simple; denoting by $N_\sigma$ the average number of
particles lying within distance $\sigma$ from a given particle during
the simulation, one simply has:
\begin{eqnarray}
u_{ex}(\beta) = {{1}\over{2}} \varepsilon N_{\sigma}(\beta). 
\label{simple}
\end{eqnarray}

If one envisions a two dimensional $\eta$-$t$ plane, then the $\eta$-route
corresponds to a horizontal path and the $t$-route to a vertical path
along this plane. If neither of the two paths crosses any phase
boundaries along its way from its starting point to its end, then
the values obtained for the excess free energy using either route
should be {\it identical}. If, on the other hand, one (or more) 
phase boundaries are encountered along the way, then differences
will occur. We have, therefore, performed simulations for various
different temperatures and density ranges to check this agreement
and to use the results as a first diagnostic tool for possible 
phase transformations on the system. Results for temperatures
$t = 0.1$, $0.2$ and $1.0$ are shown in Figs.\ \ref{simul}(a), (b)
and (c) respectively. As can be seen, the two routes yield identical
results (within `experimental' errors) for the highest temperature,
up to $\eta = 1.8$. However, for the two lowest temperatures, 
discrepancies start to appear, for $t = 0.1$ at about $\eta = 0.5$
and for $t = 0.2$ at about $\eta = 0.7$. As this is a clear indication
of a phase transition located in the neighborhood of these $\eta$-values,
neither the $\eta$- nor the $t$-route results can be considered as
reliable estimates of the free energy of the system for
values of $\eta$ exceeding the above. However, they 
can be used in conjunction with our theoretical results for the 
free energy of the {\it crystal} phase in order to draw some general
conclusions regarding the {\it topology} of the phase diagram, on the
one hand, and to trace it out in more detail on the other. These 
considerations are
presented in the section IV.

\section{The solid phases}

\subsection{General considerations}

In the solid phase the one-particle density $\rho({\bf r})$ is 
position-dependent, a property that characterizes the crystal as an
inhomogeneous phase. In the last twenty years, a common theoretical
tool which provides for a satisfactory treatment of the freezing
transition has been density-functional theory (DFT). In DFT, 
the crystal is viewed as a spatially inhomogeneous fluid and 
the properties of the homogeneous phase are used to evaluate the
free energy of a candidate crystalline structure, for a review
see \cite{singh,loewen}. Among the most popular versions of DFT is
the modified weighted density approximation (MWDA) of Denton and
Ashcroft \cite{mwda}, which has been proven to be quite reliable 
for the case of the hard-sphere freezing transition. 

In common applications of the MWDA, the one-particle density of
the candidate crystal structure is modeled as a sum of 
normalized Gaussians
centered around lattice sites, and the width (localization) of the
Gaussians is used as a variational parameter until a minimum of
the free energy is found. Typically, one makes in the MWDA the
assumption that there is just one particle per lattice site.
This is manifested in
in the usual parametrization for the one particle density 
mentioned above which reads as:
\begin{eqnarray}
\rho({\bf r}) = \Biggl({{\alpha}\over{\pi}}\Biggr)^{3/2}
                \sum_{\{{\bf R}\}}\exp[-\alpha ({\bf r} - {\bf R})^2],
\label{single}
\end{eqnarray}
where $\{{\bf R}\}$ denotes the set of Bravais lattice vectors and
$\alpha$ is the localization parameter. We call the version of the 
MWDA where the above assumption is made the constrained-MWDA. 

In principle, one would like to have at hand the possibility of
treating the average site occupancy as an {\it additional} 
parameter in the theory. Then, the restricted  
parametrization (\ref{single}) above, 
must be replaced by the more general expression:
\begin{eqnarray}
\rho({\bf r}) = x \cdot \Biggl({{\alpha}\over{\pi}}\Biggr)^{3/2}
                \sum_{\{{\bf R}\}}\exp[-\alpha ({\bf r} - {\bf R})^2],
\label{general}
\end{eqnarray}
where $x$ stands for the average site occupancy and is to be treated
as a variational quantity. For a HS-crystal (and in general for
all diverging potentials which do not allow multiple occupancy),
it is natural to expect $x \leq 1$. For the interaction at hand,
this general parametrization is quite essential, if DFT is to be 
used, for the following reason: as the density of the crystal is
increased beyond the close-packing limit of the considered 
structure, it is expected that it will be favorable for the system to
form fractions of pairs, triplets etc. Indeed, whereas for a
crystal with single occupancy the energy cost per site above the
close packing limit is equal to one-half of the number of nearest 
neighbors, formation of a number of pairs brings about a much lower
cost, simply equal to the number of paired sites. At the same time,
by pairing the lattice constant `opens up' and overlaps between
nearest neighbors are avoided. The tendency for the formation of
of composite particles, or `clusters' is also 
manifested already in the fluid, through the dramatic
increase of the liquid-state $g(r)$ towards the origin mentioned
in the previous section.

The difficulty we are faced with, however, is that a free minimization
of the MWDA-functional does {\it not} yield a physically acceptable
value for $x$ for the case of hard spheres. Indeed, it has
been found \cite{ohnesorge} that the minimum of the unconstrained
MWDA occurs for a site occupancy $x = 1.31$, an obvious physical impossibility
for hard spheres. It follows then that the results of a free
minimization of the MWDA-functional cannot be trusted, at any
temperature.
If, one the other hand, the general parametrization
given by Eq.\ (\ref{general}) above is maintained, but the domain
of acceptable solution for $x$ is restricted by hand to $0 \leq x \leq 1$,
then the value $x = 1$ is obtained as the minimum. Hence,
the {\it constrained} MWDA gives quite reliable 
results for the entropic free energy of a HS crystal. 

Clearly, the possibility of clustering appears as a mechanism for the
lowering of the free energy of the crystal mainly for packing
fractions exceeding the close-packing limit $\eta_{CP}$; at low
temperatures (with which we are concerned here), we can still
use the constrained MWDA for $\eta < \eta_{CP}$ and obtain information
about the structure of solids with single occupancy. We carried out
the MWDA calculation for temperatures $0.0 \leq t \leq 0.3$, using
the PY-results as input for the fluid structure and free energy.
The advantage of the 
MWDA is that the solid is mapped onto an effective liquid having
a `weighted packing fraction' $\hat \eta$ which is much lower than
$\eta$ of the solid \cite{mwda}, typically $\hat\eta \approx 0.30$.
The necessary ingredients for the MWDA are the values of the
structure factor $S({|{\bf K}|; \hat \eta})$ of the liquid at
the nonzero reciprocal lattice vectors ${\bf K}$ of the crystal
and the free energy per particle of the fluid again at packing
fraction $\hat \eta$ \cite{mwda}.  
For such low packings, the PY-solution is reliable, as can
be seen from Fig.\ \ref{compsk} for the structure factor and from 
Figs.\ \ref{simul}(a), (b), where we show the free energy
curves obtained from the compressibility route of the PY-solution,
demonstrating that they run very close to the simulation results
for low packings. We found that for all temperatures, the solid
free energies were indistinguishable from the HS ($t = 0$) result,
demonstrating that the structure of the solids below close packing,
even
at finite temperatures is identical to the HS-solid, i.e.\ the
particles avoid any overlap. We will make use of this preliminary 
result shortly. However, for the study crystals with 
packing fraction exceeding $\eta_{CP}$, the MWDA is unsuitable 
for the reasons
explained above, and we have to resort to a different approach.

\subsection{A cell model for the clustered solids} 

Let us consider, to begin with, a HS-solid of $N$-particles 
(HS diameter $\sigma$, mass $m$) enclosed
in volume $\Omega$, having packing fraction $\eta$,
and site occupancy equal to unity. The partition function $Q_N(\eta)$
is given by:
\begin{eqnarray}
Q_N(\eta) &=&  
    \Biggl({{4\pi}\over{h^3}} 
    \int p^2 e^{-\beta p^2/(2 m)} dp\Biggr)^N \times
    {{1}\over{N!}}\int_{\Omega} d{\bf r}_1 d{\bf r}_2 \cdots d{\bf r}_N
    e^{-\beta V({\bf r}_1, {\bf r}_2, \cdots {\bf r}_N)} \cr
    \cr
    &\equiv& \Theta_N \times Z_N(\eta),
\label{hspart}
\end{eqnarray} 
where $h$ is Planck's constant,
$\Theta_N$ is the kinetic and $Z_N$ 
the configurational part of the partition 
function. For the evaluation of the latter, we adopt the cell 
model \cite{cell1,cell2,cell3,cell4} which exploits the picture 
of particle in a solid as being confined in cells of cages formed
by the neighboring ones
from which it cannot escape. We emphasize here that we employ
the cell model only as an intermediate step in order to establish
a relation between the free energy of a clustered crystal and
that of a HS crystal and not as a computational tool in order to
actually calculate these quantities.
The packing fraction $\eta$ and the candidate crystal structure
determine the volume of the cell, also called free volume $v_f(\eta)$.
Then,
the particles in the solid can be treated as distinguishable. Since
within the cell the Boltzmann factor is unity, the
configurational partition function is given by:
\begin{eqnarray}
Z_N(\eta) = \Biggl(\int_{v_f(\eta)} d{\bf r}\Biggr)^N = v_f^N(\eta).
\label{hscell}
\end{eqnarray}
Strictly speaking, the expression above provides
only a lower bound to the true partition function of the 
crystal \cite{matthias,schmidt}.

Combining Eqs. (\ref{hspart}), (\ref{hscell}) above, we obtain the 
free energy per particle of a HS-crystal having packing fraction $\eta$
and site occupancy one, as:
\begin{eqnarray}
{{\beta F_{\rm HS}(\eta)}\over{N}} \equiv f_0(\eta) = 
 - \ln\Biggl[{{v_f(\eta)}\over{\Lambda}^3}\Biggr],
\label{f0}
\end{eqnarray}
where $\Lambda \equiv (2 \pi m k_B T/h^2)^{1/2}$ is the thermal de Broglie
wavelength.

Let us now proceed in an analogous way for the general case $t \ne 0$.
As mentioned above, we expect the formation of doublets, triplets etc.\
in the crystal. Clearly, as the density is increased, more and more
complicated composites will appear (quadruplets, quintuplets etc.) 
To keep the discussion simple 
(and the theory computationally manageable) we
restrict ourselves here to clusters up to triplets only. 

Let us then consider $N$ particles in a crystal with $N_s$ sites.
Of these $N_s$ sites, $N_1$ are occupied by a single particle, $N_2$
by pairs and $N_3$ by triplets. We set:
\begin{eqnarray}
{{N_1}\over{N_s}} \equiv s, \;\;\;\;\;
{{N_2}\over{N_s}} \equiv z \;\; {\rm and} \;\
{{N_3}\over{N_s}} \equiv w.
\label{zw}
\end{eqnarray}
Clearly, $s+z+w = 1$ and $N = (1 + z + 2 w)N_s$.
The result of the formation
of composites is that the `clustered solid' has a lattice constant 
which corresponds not to $\eta$ but to a new, 
{\it effective}
packing fraction $\gamma$ which is lower and related to $\eta$ by:
\begin{eqnarray}
\gamma = {{\eta}\over{1 + z + 2 w}}.
\label{effeta}
\end{eqnarray}

The idea is that the system will find it favorable to create as many 
clusters as possible so as to bring about an effective packing 
$\gamma$ which is below $\eta_{CP}$. This way, the energy 
cost comes entirely from the sphere overlaps
in the clusters themselves; otherwise, the lattice cell is now
large enough, so that the expensive, multiple overlaps with the
neighbors are avoided. This assumption is corroborated by the 
MWDA-results for the single-occupied solids below $\eta_{CP}$. 
Indeed, it was found that, for low temperatures and $\eta < \eta_{CP}$,
the system behaves essentially as a HS-crystal. Hence, our model
for the clustered solid is the following: enough clusters are
formed so that the effective packing fraction $\gamma$ is always 
below $\eta_{CP}$ and, once this has been achieved, 
each object occupying a lattice site (being a single particle
or a composite) acts as a hard sphere with respect to any other 
object occupying a neighboring site. 

With these assumptions in mind, we now proceed with a cell model
for the clustered solid. The free volume $v_f$ is now dictated by
the packing fraction $\gamma$.
Each site occupied by a pair brings about an energy cost $\varepsilon$
and each site occupied by a triplet a cost $3 \varepsilon$. Taking 
into account the indistinguishability of the particles in the 
clustered sites,
we can now write down an expression the partition
function of our clustered crystal which, at this stage, does {\it not}
include the entropy of mixing:
\begin{eqnarray}
Q_N(\eta,t) = \Theta_N\times
               \Biggl[\int_{v_f(\gamma)} d{\bf r}\Biggr]^{N_1}\times
 \Biggl[{{e^{-\beta\varepsilon}}\over{2!}}\int_{v_f(\gamma)} d{\bf r}
                \int_{v_f(\gamma)} d{\bf s}\Biggr]^{N_2}\times\cr
 \cr
 \Biggl[{{e^{-3\beta\varepsilon}}\over{3!}}\int_{v_f(\gamma)} d{\bf r}
       \int_{v_f(\gamma)} d{\bf s} \int_{v_f(\gamma)} d{\bf t}
       \Biggr]^{N_3}.\qquad\qquad
\label{cell}
\end{eqnarray}
Using the relation $N = N_1 + 2N_2 + 3N_3$, performing the volume
integrals above and taking the logarithm, we obtain:
\begin{eqnarray}
-{{\ln Q_N(\eta,t)}\over{N}} = 
-\ln\Biggl[{{v_f(\gamma)}\over{\Lambda^3}}\Biggr] +  
                 \Biggl({{N_2 + 3 N_3}\over{N}}\Biggr)t^{-1} + 
                 {{N_2}\over{N}} \ln 2 + {{N_3}\over{N}} \ln 6.
\label{logar}
\end{eqnarray}
The first term is, according to Eq.\ (\ref{f0}),
nothing else but $f_0(\gamma)$, the free energy of a HS-crystal
having packing fraction $\gamma$.
The above expression is not yet the free energy of the clustered crystal
as it does not include the `mixing-entropy' contributions arising from
all the possible ways of choosing the $N_2$ and $N_3$ sites which are
occupied by clusters. This mixing entropy is simply:
\begin{eqnarray}
S_{mix} = k_B \ln W,
\label{mixentr}
\end{eqnarray}
where $W$ is precisely the number of ways of choosing $N_2$ and $N_3$
sites out
of $N_s$ for the multiple occupancies. It is straightforward to show that:
\begin{eqnarray}
W = {{N_s!}\over{N_2!N_3!(N_s-N_2-N_3)!}}.
\label{combinat}
\end{eqnarray} 

Finally, the $z$- and $w$-dependent {\it variational} expression for
the free energy per particle of a solid with clusters is given by:
\begin{eqnarray}
{{\beta\tilde F(\eta,t;z,w)}\over{N}} \equiv \tilde f(\eta, t; z, w) =
-{{\ln Q_N(\eta, t)}\over{N}} - {{S_{mix}}\over{k_B N}}.
\label{sum}
\end{eqnarray}
Collecting the results from Eqs.\ (\ref{zw})-(\ref{logar}) and
(\ref{combinat}) above, we finally obtain:
\begin{eqnarray}
\tilde f(\eta, t; z, w) = 
f_0\Biggl({{\eta}\over{1+z+2w}}\Biggr) + 
   \Biggl({{z+3w}\over{1+z+2w}}\Biggr) t^{-1} + 
   \Biggl({{z\ln 2 + w \ln 6}\over{1 + z + 2w}}\Biggr) + \cr
   \cr
   {{1}\over{1+z+2w}}[z\ln z + w \ln w + (1-z-w) \ln (1-z-w)].
\label{variat}
\end{eqnarray}
The quantities $z$ and $w$ are variational parameters, 
as there are no chemical potentials controlling the site
occupancy, hence the free
energy per particle of the solid is given by:
\begin{eqnarray}
f(\eta, t) = \min_{\{z,w\}} \tilde f(\eta, t; z,w).
\label{minimize}
\end{eqnarray}

In our considerations we have examined both the fcc- and bcc-solids,
finding that the fcc is favorable always. So we restrict the discussion
to this structure only. For the fcc, $\eta_{CP} = 0.74$.
For the free energy
per particle of the HS-solid, $f_0$, at packing fraction 
$\gamma = \eta/(1+z+2w) < \eta_{CP}$
we use the results from the constrained-MWDA.
An important result from the MWDA is that the 
fcc HS-solid is {\it mechanically unstable} below $\eta = 0.46$, 
i.e.\ the MWDA-free energy cannot be minimized by a nonzero value
of $\alpha$ if the packing fraction is below the value mentioned
above. We have, thus, imposed an artificially high (practically
infinite) value for the function $f_0(\eta)$ for $\eta < 0.46$ and
proceeded with the numerical minimization. The latter must be
performed in the triangular domain which is enclosed in the $z$-$w$ plane by
the boundaries: $0 \leq z \leq 1$; $0 \leq w \leq 1$; and
$z + w \leq 1$.

\subsection{Comparison with simulations}

In order to check the reliability of the fraction of doubly 
occupied lattice sites $z$ as obtained from the above described 
theoretical model, we performed a numerical
calculation of the free energy $\tilde F$ of the fcc solid at fixed 
temperature $t=0.1$ and fixed particle volume fraction $\eta=0.8$, 
where the theory predicts $w=0$, i.e.\ there are only singlets and 
doublets in the crystal. For that purpose,
we took advantage of a thermodynamic integration
method initially introduced by Frenkel and Ladd
\cite{Frenladd,Frenkel}. In this Monte Carlo method,
the free energy of the investigated system is calculated by transforming
the system reversibly
into a harmonic Einstein crystal of the same crystal symmetry,
whose free energy $F^{Ein}$ is known analytically.
The crystal symmetry of the reference crystal is
simply characterized by the zero temperature lattice sites of the
$N$ simulated particles 
${\{{\bf R}\}}_0^N=
({\bf R}_{0,1},$ $\ldots, {\bf R}_{0,N})$.
A throughout extensive description of the method can be found in
Ref.\ \onlinecite{Frensmit}.

In our specific use of this method, we choose the lattice sites
${\bf R}_0^N$ of the
reference harmonic crystal to be partially doubly occupied, i.e.\
we set ${\bf R}_{0,i}={\bf R}_{0,j}$ for some randomly chosen particle
numbers $i$ and $j$. We do not allow three particles to have the same
reference crystal position. So, the reference crystal structure is
characterized by its crystal symmetry (chosen to be fcc in our case),
its particle volume fraction $\eta$, and its fraction of doubly
occupied lattice sites $z$. For fixed $\eta$ and $z$, $\tilde F$ could then 
be calculated as described in detail in Refs.\
\onlinecite{Frenkel,Frensmit}.
We performed calculations for various
pairing fractions $z$, ranging from $0.35$ to $0.80$, fixing the
temperature at $t=0.1$ and the density at $\eta=0.8$. In all simulations,
the number of particles was between 500 and 700, therefore
finite size effects could be neglected.

Since our Monte Carlo simulations were always performed for
one specific realization of the singlet-and-doublet fcc solid,
we had to add the mixing entropy $S_{mix}$
[as given by Eqs.\ (\ref{mixentr}) and (\ref{combinat})]
to our Monte Carlo free energy results.
In principle, in the Monte Carlo simulations,
the system was free to explore the configuration space associated with
the various possible realizations of the fcc solid,
since we did not restrict the particle coordinates to distinct regions
in the simulation box. However, this would have required very long simulation
runs since very large mean-square displacements of the particles would
have been needed. Since in our simulations the mean-square displacements of
the particles were in the order of the lattice spacings, we had to
take into account the mixing entropy $S_{mix}$.

In Fig.\ \ref{pairs}
we show our Monte Carlo results for the free energy of the
fcc solid with singly- and doubly-occupied sites
including the mixing entropy $S_{mix}$, as a function of
$z$ for fixed $t$ and $\eta$. Also shown is the corresponding results of our
above described theoretical model [i.e.\ from Eq.\ (\ref{variat}) with
$w = 0$.] Obviously, the agreement of
the pairing fractions $z_{min}$, where the free energy of the fcc solid
achieves a minimum is very good. We have also done the same check at
different $\eta$-values, obtaining similar agreement; thus, the 
theory described above provides a reliable method for the calculation
of the free energy of the crystals.

\subsection{Results of the variational calculation}

We now present in detail the results obtained from the theory in the 
range of thermodynamic parameters $0 \leq t \leq 0.3$ and $\eta \leq 2.2$.
First, we introduce a terminology to characterize the various types
of fcc-solids with respect to the fractions of sites occupied by
clusters, as follows:\\
\noindent
(i) S-solid if $s = 1$, $z = w = 0$;\\
\noindent
(ii) SP-solid if $0 < s < 1$, $0 < z < 1$ and $w = 0$;\\
\noindent
(iii) P-solid if $s = 0$, $z = 1$ and $w = 0$;\\
\noindent
(iv) PT-solid if $s = 0$, $0 < z < 1$ and $0 < w < 1$;\\
\noindent
(v) SPT-solid if $0 < s < 1$, $0 < z <1$ and $0 < w < 1$, and\\
\noindent
(vi) T-solid if $s=z=0$, $w = 1$.

These are the six types of solids that come out of the minimization. 
In Fig.\ \ref{szw} we show the dependence of $s$, $z$ and $w$ on
the fcc-packing fraction $\eta$ for $t = 0.05$. The
typical scenario that materializes, at least for temperatures 
$t {< \atop \sim} 0.1$ is the following: for packing fractions 
$\eta {< \atop \sim} \eta_{CP}$, we have the usual S-solid, as there is
no particular gain for clusters to be formed. At higher densities,
pairs start to appear and an SP-solid is formed. The pair fraction
grows with density at the expense of the
singly-occupied sites. Depending on the temperature, the fraction of
pairs may reach the value unity at about $\eta \approx 2\eta_{CP}$
before any triplets appear, thus forming a P-crystal; this happens
for $t {< \atop \sim} 0.05$. 
For higher temperatures, triplets 
appear while both $s$ and $z$ are nonzero, thus giving rise to a
SPT-solid. By further increase of the density, the single-occupancy
sites disappear altogether and a PT-solid emerges. Then, the pairs
start being replaced by triplets completely and a T-solid takes the
place of the PT-solid.

As shown in Fig.\ \ref{szw}, the fractions of multiply occupied sites
approach zero in a continuous way. Thus, we are having a sequence of
{\it second-order clustering transitions} in the solid which gets more
and more complicated as the packing fraction grows. 
Whether all this sequence of transitions will
actually appear in the phase diagram depends also on the competition
with the liquid free energy.
The full phase diagram,
including the freezing transition, is discussed
in the following section.

\section{The phase diagram} 

In this section we determine the low-temperature phase diagram of the
system, putting together the results obtained for the free energy of
the solid, obtained by the procedure described previously, and those
for the fluid free energy coming from the simulations. A representative
case for $t=0.1$ is shown in Fig.\ \ref{comtan}. The first question 
to be addressed is the topology of the phase diagram, in particular
the possibility of the existence of {\it reentrant melting}, i.e.\
a remelting of the solid at higher densities. This is a realistic
possibility which in fact materializes for the bounded
Gaussian potential of Stillinger \cite{still,stillweb,stillrev}. 

Referring to Fig.\ \ref{comtan}, we see that if the $\eta$-route result
for the is taken as the `true' liquid free energy, then we would have
indeed reentrant melting; in fact, for this temperature the solid 
would be marginally stable at $\eta \approx 1.0$, i.e.\ $t=0.1$ would
be very close to a `maximum freezing temperature' above which no 
thermodynamically stable solids would exist. However, were this to
be the case, then the $t$-route to the fluid free energy would have
crossed no phase boundaries along its way. Thus, the $t$-route would
have given the true fluid free energy at high packings and this, in turn,
ought to lie below the solid free energy, consistently with the 
reentrant-melting scenario. Obviously, this is not the case. This
leads us to the conclusion that there is no reentrant melting, at least
not in the range of densities considered here. Instead, there is freezing
into a fcc-solid, followed by a cascade of clustering transitions as
described in the previous section. 
 
The coexistence densities for the freezing transition are determined
by performing a common-tangent construction on the fluid- and 
solid-free energy curves. As mentioned previously, none of the
$\eta$- or $t$-route curves can be considered as the `true' free
energy of the fluid beyond the point where they start to diverge
from each other. However, the $\eta$-route curve is in a way 
`more wrong' than the $t$-route curve in the sense that it yields,
at high densities, fluid free energies which are lower than their
solid counterparts and this leads to the contradiction explained
above. Thus, the correct free energy of the fluid must follow
a curve which is identical to the simulation results up to the 
point where the two routes agree (and where the liquid is stable) and
then it must cross the solid free energy and run above it (and
thus the liquid is there metastable). In this sense, the fluid
free energy is `closer' to that obtained by the $t$-route than 
the one obtained from the $\eta$-route. Therefore, we have performed
the common-tangent construction using the $t$-route result for
the fluid. As the lower end of the common tangent ends up lying
in the region where the $t$-route results are indeed reliable,
the precise shape of the liquid free energy curve for densities
{\it beyond} freezing is immaterial.

>From the more quantitative point of view, 
the fact that the coexistence region lies {\it precisely} in the domain 
where the $\eta$- and $t$-routes yield results that begin to diverge
is an independent
confirmation for the theoretical approach we employed for the solid.
Indeed, this discrepancy is the signature of a phase transition 
which now comes about to be located in the right place by means of
a completely independent theoretical approach for the crystal.
The same agreement was found at all temperatures we considered.

Putting now everything together, we trace out the liquid-solid 
coexistence curves as well as the boundaries of the second-order
transitions between the crystals with the different types of 
clustering. The phase diagram obtained in this way is shown in
Fig.\ \ref{phdg}. The region of stability of the T-phase is
artificially enlarged. The reason is that, in order to determine
with accuracy the stability for a given type of clusters, at least
the next type of cluster must be put into the theory, i.e.\ 
quadruplets for the T-solid etc. As this is an increasingly 
complicated procedure, we have not done this here. However, in
view of the results already obtained, we expect that the
solid will proceed with more and more clustering at increasing
density, thus giving rise to a quite interesting phase diagram.

\section{Discussion and conclusions}

We have considered a toy model of penetrable spheres characterized
by an interaction which imposes a constant energy cost if there 
is any overlap between the spheres (no matter how strong) and zero
otherwise. Although the model is quite simple, the form of the
interaction, which favors full overlaps between the particles,
brings about quite a few interesting features. As a first remark,
we have found an inadequacy of the traditional liquid-state 
integral equation theories to describe in a satisfactory way
the high-density fluid phase of the system. We believe that this
shortcoming can be traced back to the inaccuracies in
the estimation of the bridge function, inherent in all approximate
closures. Such inaccuracies are not dramatic if we are dealing with
a unbounded interaction. In
those cases, the different closures give results which differ on
the amount of structure of, say, the radial distribution function
$g(r)$ outside some effective core where $g(r)$ vanishes. However,
since the bridge function $B(r)$ attains its highest values {\it precisely}  
for $r \to 0$, if the bare interaction is not strong enough to 
dominate over the bridge function, then inaccuracies in the 
determination of the latter become really crucial. Thus, it is
not surprising that in our case the problem becomes more severe
as the density grows (because then $B(r)$ grows as well) and/or
as the temperature is raised (because then the bare interaction
$\beta \phi(r)$ diminishes.) 

To the best of our knowledge, the only other bounded interaction
for which an attempt has been made to trace out the phase diagram 
is the Gaussian model of Stillinger \cite{still,stillweb,stillrev}.
In that case, it was found that the model displayed reentrant
melting. In Ref.\ \onlinecite{still}, 
some general criteria for the mathematical form of
the interaction were laid down and it was stated that for any
pair potential meeting those criteria, reentrant melting 
behavior had to be expected. These conditions are: (i) the 
interaction must be bounded at the origin; (ii) it must vanish
strongly enough at infinity to be integrable and (iii) it must be
differentiable at least four times. Our interaction satisfies these
requirements, with the exception of (iii) since it has a singularity
at $r = \sigma$ and it is not differentiable there. However, this
does not constitute a serious violation as one could easily imagine
an analytic potential that would run arbitrarily close to our
`step function' and for that potential the results would be 
practically identical to the ones found here.
However, another important 
ingredient that goes into reaching these general conclusions is the
assumption
that the solid (or solids of different crystal symmetry) which are
`nested' between the fluid at low- and high-densities have single
site occupancy. We have not found reentrant melting in our case,
at least for the range of densities and temperatures we considered.
Although we cannot exclude this possibility at some other region
of the phase diagram, we believe that the arguments of 
Ref.\ \onlinecite{still} do not apply to our case, precisely due to 
the clustering in the solid which takes place in our model.
For the same reasons, our results are at odds with those of
Marquest and Witten \cite{marquest} who found regions of stability
of the bcc- and simple-cubic structures at growing density, based 
on calculations of the ground-state energy, making the assumption
of single occupancy in the crystal. We find instead that a cascade of
second-order transitions takes place in the crystal.

\section*{ACKNOWLEDGMENTS} We are pleased to acknowledge useful discussions
with Prof. D. Frenkel and Dr. A. R. Denton. M. W. thanks the Deutsche 
Forschungsgemeinschaft for support within the SFB 237.

\begin{figure}
\caption[dum01] {Comparison of the radial distribution function $g(r)$
as obtained from simulation, and the PY- and HNC-closures, for  
a system of penetrable spheres at reduced temperature $t=0.2$
and packing fraction $\eta = 0.3$.} 

\label{compgr}
\end{figure}

\begin{figure}
\caption[dum02] {Comparison between the simulation result and the
PY-closure for the structure factor $S(k)$ at the same point as
in Fig.\ \ref{compgr}.}

\label{compsk}
\end{figure}

\begin{figure}
\caption[dum03] {Comparison between the simulation result and the
PY-closure for the function $g(r)$ at $\eta=0.5$ and $t=0.2$. Note
the dramatic increase of $g(r)$ from simulation inside the core.
The simulation value for $g(r)$ at $r=0$ is in fact equal to 18.5.}

\label{comppy}
\end{figure}

\begin{figure}
\caption[dum04] {Free energy densities as obtained by the $\eta$- and
$t$-routes of the simulation. (a) $t=0.1$; (b) $t=0.2$; (c) $t=1.0$.
The solid lines in (a) and (b) denote the results obtained by 
using the compressibility route of the PY-solution and demonstrate
that for low densities the PY-closure gives reasonable results for
this quantity.}

\label{simul}
\end{figure}

\begin{figure}
\caption[dum05] {The variational free energy of an fcc-solid
having packing fraction $\eta=0.8$ at temperature $t = 0.1$ as
a function of the fraction of sites occupied by pairs.}

\label{pairs}
\end{figure} 

\begin{figure}
\caption[dum06] {The fraction of sites with single,
double and triple occupancy as a function of $\eta$ for fcc-solids at
temperature $t=0.05$.}

\label{szw}
\end{figure}

\begin{figure}
\caption[dum07] {Free energy densities for the fluid, as obtained by
using the $\eta$- and $t$-routes in the simulation and for the solid
as a result of the theory. The reduced temperature is $t=0.1$.}

\label{comtan}
\end{figure}

\begin{figure}
\caption[dum08] {The phase diagram of the penetrable sphere model.
The thick lines denote the first-order freezing transition and the 
shaded region is the liquid-solid coexistence region. The dashed lines
denote second-order clustering transitions in the solid. As explained 
in the text, the region of stability of the T-solid is artificially 
enlarged due to the lack of the possibility of formation of 
four-particle clusters in our theory.}

\label{phdg}
\end{figure} 

\end{document}